\documentclass[12pt]{article}
\usepackage[latin1]{inputenc}
\usepackage{amssymb}
\usepackage{graphicx}
\usepackage{float}
\newcommand{\text}{\rm}

\newcommand{\bb}{\begin{equation}}
\newcommand{\ee}{\end{equation}}
\newcommand{\bega}{\begin{eqnarray}}
\newcommand{\ega}{\end{eqnarray}}
\newcommand{\begae}{\begin{eqnarray*}}
\newcommand{\egae}{\end{eqnarray*}}

\newcommand{\h}{\hspace*{4ex}}

\newcommand{\cent}{\centerline}
\newcommand{\vs}{\vspace*}

\begin{document}

\baselineskip 0.8cm

\begin{center}

{\large {\bf Modelling the spatial shape of nondiffracting beams: experimental generation of Frozen Waves via
holographic method} $^{\: (\dag)}$ } \footnotetext{$^{\: (\dag)}$ E-mail addresses for contacts:
mzamboni@dmo.fee.unicamp.br}


\end{center}

\vs{5mm}

\cent{T\'{a}rcio A. Vieira$^{\: (1)}$, Michel Zamboni-Rached$^{\: (2)}$ and Marcos R.R. Gesualdi$^{\: (3)}$}

\vs{0.2 cm}

\centerline{{\em $^{\: (1,3)}$Universidade Federal do ABC, Santo André, SP, Brazil.}}

\vs{0.2 cm}

\centerline{{\em $^{\: (2)}$University of Campinas, Campinas, SP, Brazil.}}

\vs{0.5 cm}

{\bf Abstract  \ --} \ In this paper we implement experimentally
the spatial shape modelling of nondiffracting optical beams via
computer generated holograms on spatial light modulators. The
results reported here are the experimental confirmation of the so
called Frozen Wave method, developed few years ago. Optical beams
of this type can possess potential applications in optical
tweezers, medicine, atom guiding, remote sensing, etc..


\section{Introduction}

\h Few years ago, in a series of papers \cite{Zamboni;FW,Zamboni-Rached:05,Zamboni-Rached:06,Zamboni-Rached:10},
an interesting theoretical method was developed, capable to furnish nondiffracting beams whose longitudinal
intensity shape can be freely chosen a priori.

\h This approach is based on suitable superposition of equal frequency and co-propagating Bessel beams, and the
resulting wave fields are called Frozen Waves\footnote{This approach is also called \emph{Frozen Wave method}}
(FWs). Besides a strong control on the longitudinal intensity pattern, this method also allows a certain control
on the transverse shape of the resulting beam.

\h Due to their unique characteristics, i.e., their nondiffracting
and spatial modelling properties, the FWs are quite interesting
for many applications such as optical tweezers, remote sensing,
atom guiding, medical purposes, etc. \cite{Zwick:09,meyers:061126,PhysRevLett.57.314}

\h Very recently \cite{Vieira:12} the FW method was experimentally
verified through the experimental generation by holographic method
of few FWs previously chosen.

\h In this paper we shall present the experimental generation of several new and very interesting FWs through
the implementation of amplitude computer generated holograms (CGHs) in spatial light modulators (SLMs). Our
results confirm, once more, the theoretical predictions of the method developed in
\cite{Zamboni;FW,Zamboni-Rached:05}, and open exciting possibilities on the applicability of these very especial
beams.

\h In the next section we make a synthesis of the theoretical FW
method. After this, in section 3, we show the experimental results
concerning to the generation of several nondiffracting beams whose
spatial shape were previously chosen. The experimental generation
is made by amplitude computer generated holograms implemented in
two types of spatial light modulators, transmission and
reflective.

\section{Summarizing the theoretical Frozen Wave method}

\h The theory of FWs was formulated in \cite{Zamboni;FW} and further improved in \cite{Zamboni-Rached:05,Zamboni-Rached:06,Zamboni-Rached:10}.

\h Here we shall summarize the method without entering into the mathematical details, which can be found in the
references above.

\h Trying to be brief, what we wish is to construct exact solutions of the wave equation representing
nondiffracting beams whose the longitudinal intensity pattern, $|F(z)|^2$, in the interval $0 \leq z \leq L$,
can be freely chosen a priori.

\h This can be done by considering a superposition of equal frequency and co-propagating Bessel beams of order
$\nu$:

  \begin{equation} \label{campo_FW}
        \Psi(\rho,\phi,z,t) = e^{-i\omega t}\sum^{N}_{n=-N}A_{n}J_{\nu}(k_{\rho n}\rho)e^{ik_{zn}z}e^{i\nu\phi}
        \end{equation}
with
        \begin{equation} \label{krho}
        k^{2}_{\rho n} = k^{2} - k^{2}_{zn}
        \end{equation}
where $k$, $k_{\rho n}$ and $k_{z n}$, are the total, the transverse and the longitudinal wave numbers
respectively of the nth Bessel beam in the superposition \ref{campo_FW}.

\h In expression (\ref{campo_FW}) the following choice is made:

        \begin{equation} \label{kz}
        k_{zn} = Q + 2\pi n /L
        \end{equation}
where $Q$ is a constant such that

\begin{equation} \label{Q}
        0 \leq Q + (2\pi/L)n \leq \omega/c
        \end{equation}
for $-N \leq n \leq N$.

\h The condition given by (\ref{Q}) ensures forward propagation only, with no evanescent waves. The constant
parameter $Q$ can be freely chosen, provided that (\ref{Q}) be obeyed, and it plays an important role
in determining the spot size of the resulting beam.

\h Still considering Eq.(\ref{campo_FW}), we adopt the following choices for the coefficients $A_n$:

        \begin{equation} \label{An}
        A_{n} = \frac{1}{L} \int^{L}_{0}F(z) e^{-i\frac{2\pi}{L}nz} dz
        \end{equation}
where, as we said, $|F(z)|^{2}$ is the desired longitudinal intensity pattern in the interval $0 \leq z \leq L$.

\h Now, it is important to notice that this longitudinal intensity pattern can be concentrated, as we wish, over
the propagation axis ($\rho = 0$) or over a cylindrical surface.

\h In the case we wish this intensity concentrated over the propagation axis, $\rho = 0$, zero order Bessel
beams (i.e. $\nu=0$) are to be used in the fundamental superposition (\ref{campo_FW}). It is also possible to
choose the spot radius, $\Delta \rho_0$, of the resulting beam by making $Q = (\omega^{2}/c^{2} -
2.4^{2}/\Delta \rho^{2}_{0})^{1/2}$.

\h Now, if we wish this intensity configuration concentrated over a cylindrical surface, so higher order Bessel
beams, i.e. $\nu \geq 1$, are to be used in (\ref{campo_FW}). In this case, the radius $\rho_0$ of the
cylindrical surface can be approximately chosen if we pick up the value of $Q$ as given by

        \begin{equation} \label{raio cilindro}
        \left.\left[\frac{d}{d\rho}J_{\nu}\left(\rho\sqrt{\omega^{2}/c^{2}-Q^{2}}\right)\right]\right|_{\rho=\rho_{0}} = 0
        \end{equation}

\section{Experimental generation of Frozen Waves via holographic method}

\h The use of spatial light modulators devices in holographic setups has became possible interesting
applications in image phase correction, code signal encrypted and generation of optical beams \cite{Vasara,Bouchal:02,Arrizon:03,Arrizon:05,Vieira:12,Remenyi:03,Nicolas:02}.

\h The experiments conducted by us to generate interesting types of FWs are based on the holographic method.
With the desired beams described by analytical exact solutions of the wave equation \cite{goodman}, we created an
amplitude Computer Generated Holograms (CGH), which is reconstructed by a nematic liquid crystal spatial light
modulator (LC-SLM).

\h More specifically, once we have chosen the desired beam spatial
shape (i.e., the  beam's longitudinal intensity pattern,
$|F(z)|^2$,  its spot radius or the radius of its cylindrical
form), it can be approximately described by the analytical and
exact FW solution (\ref{campo_FW},\ref{kz},\ref{Q},\ref{An}). The
amplitude CGH is constructed from the FW complex field  $\Psi
(\rho, \phi, z,t)$ (called FW-CGH) at the origin of the
propagation direction, i.e. at $ (z = 0) $, and it is given by the
following transmittance hologram equation,
\begin{equation} \label{hologram_eq}
                H(x,y)=1/2\left\{\beta(x,y) + \alpha(x,y)\cos[\phi(x,y)-2\pi(\xi x + \eta y)]\right\}
        \end{equation}

where, $\alpha(x,y)$ and $\phi(x,y)$ are amplitude and phase of
the FW complex field $\Psi (\rho, \phi, z,t)$, respectively. For
reducing the noise of the signal spectrum hologram, the
conventional bias function $\beta(x,y)=[1+\alpha^{2}(x,y)]/2$ is
taken as a soft envelope of the amplitude $\alpha(x,y)$
\cite{Arrizon:05}. To separate different diffraction orders from
the encoded complex field $\Psi(\rho,\phi,z,t)$, the off-axis
reference plane wave $\exp[i2\pi(\xi x + \eta y)$ is used. In
Fourier plane, the center of signal information is shifted to
values of spatial frequencies $(\xi,\eta)$ and should be chosen
according to difraction efficiency and bandwidth of the SLM
\cite{Arrizon:03,Vieira:12}.

\h To guarantee the efficient generation of the FW in the chosen interval, we have used (to the FW-CGH) a
circular aperture of minimum diameter $D$ given by
        \begin{equation} \label{Raio_minimo}
                D_{min} \geq 2L\left[\left(\frac{k}{k_{zn=-N}}\right)^{2}-1\right]^{1/2}
        \end{equation}

\h The parameters $Q$ and $L$ give us, via Eq.(\ref{Q}), the maximum number, $2N_{max} + 1$, of Bessel beams in
the superposition (\ref{campo_FW}). If we consider $Q>k/2$
    (as usually occurs), so
        \begin{equation}\label{Nmax}
            N_{max} = \left[ L(k-Q)/2\pi \right]
        \end{equation}
        where [.] is the floor function, i.e., $N_{max}$ is the greatest integer less than or equal to $L(k-Q)/2\pi$.

\subsection{Holographic experimental setups}

\h In this work we have experimentally generated seven different and interesting FWs, in five of them it was
used a transmission SLM, being the other two created with a reflective one. The most significant differences
between them is the pixel resolution, and consequently, bandwidth and effective display area. We will see later
the implications of this differences in holographic reconstruction processes of the FW complex field.

\begin{figure}[H]
                \centering
                \includegraphics[scale=2.3]{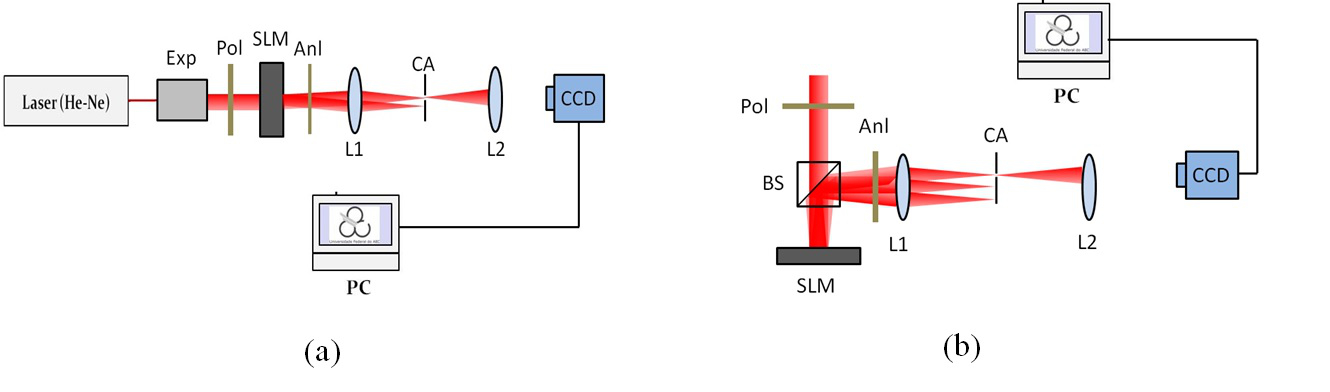}
								\caption{(a) experimental Setup 1 (b) experimental Setup 2, for FW generation,
                where SF is a spatial filter, L's are lenses, Pol is polarizer, Anl is analizer, CA is a circular aperture mask,
                CCD is the camera. In (a) SLM is a LC2002 transmission SLM and  (b) SLM
                is a LC-R1080 reflective SLM,}
								\label{arranjo}
    \end{figure}

\h In the experimental holographic setups for FW generation, Fig.\ref{arranjo}(a) to transmission SLM (Setup 1)
and Fig.\ref{arranjo}(b) to reflective SLM (Setup 2), we have a He-Ne laser ($632.8 nm$) that is expanded and
collimated (Exp) in a SLM device. Here we use the amplitude modulation with the polarizer $Pol$ (angle
0$^\circ$) and analyzer $Anl$ (angle 90$^\circ$) measured with respect to the input axis in the SLM. The 4-f
spatial filtering system is used for FW experimental generation.

\h The transmission modulator used in setup 1 is the LC-2002 SLM
model, Holoeye Photonics. It has an array of pixels $(800 \times
600)$ with each pixel measuring 32 micrometers, the shortest side
of the display possesses $19,2 mm$ and the bandwidth $\delta p=3.1
\times 10^{4}\,m^{-1}$.

\h The reflective SLM used in setup 2 is the LC-R1080 SLM model, Holoeye Photonics, that possesses each pixel
measuring $8.1 \mu m$ in a display matrix $(1920 \times 1200)$, the lowest edge with $9.7 mm$ and the bandwidth
$\delta p = 12.346\times 10^{4} m^{-1}$. Considering the limitations described above, this implies that the
limit for the diameter of the CGH is approximately two times lower and bandwidth $\delta p$ is four times higher
compared to those of Setup 1.

\h In more detail, in both setups the LC-SLM (FW-CGH) is placed at the input plane (focus of lens $L1$) and, a
spatial filtering mask ($SF$, band-pass circular pupil), at the Fourier plane, to selects and transmits the
shifted signal spectrum generating the FW field $\Psi(\rho,\phi,z)$ at the output plane of the setup. As a
result, we have the propagation of the desired FW, whose intensity is registered with a CCD camera that can be
displaced along the distance $0 \leq z \leq L$.

\subsection{Results}

\h The selection of parameters to be used in the experimental
implementation for FW generation should consider the properties of
the spatial light modulator that is being used. Two important
parameters of the SLMs are limiting in this process: the bandwidth
and the length of the shortest side of the SLM display. The first
limitation is associated with the loss of information in the
reconstruction process of the computer generated hologram (CGH);
and, the second limitation is related to the diameter that the CGH
(containing the complex field superposition) should possess in
order to guarantee the efficient generation of the FW in the
required spatial range $0 \leq z \leq L$.

We will discuss and compare the experimental results of FWs fields
generated by holographic method using SLMs.

\h In the generation of the first five FWs below it was used the experimental setup 1, and in the last two the
setup 2 was adopted. In all cases the procedure is the same: first, we choose the desired longitudinal intensity
pattern, $|F(z)|^2$, for the nondiffracting beam; after this, we use the solution (\ref{campo_FW}) with $k_{zn}$
and $A_n$ given by Eqs. (\ref{kz}) and (\ref{An}). The value of $Q$, and consequently the value of $N_{max}$,
can be chosen through the desired beam spot size, or according to the resolution limit of the SLM. In solution
(\ref{campo_FW}), we use $\nu=0$ or $\nu \geq 1$ depending if the desired longitudinal intensity pattern should
be on the $z$ axis or on a cylindrical surface, respectively.

\h The examples of FWs considered here are designed to furnish nondiffracting beams with interesting intensity
patterns. For transmition SLM we use $Q = 0.9999943\,k$, $N=9$ and $L=100\,$cm. For reflective SLM we adopt $Q =
0.99996\,k$, $N = 24$, and again $L=100\,$cm.

Unless otherwise stated, we will consider FWs generated from superpositions of zero-order Bessel beams, i.e., we
will use $\nu=0$ in the FW solution (\ref{campo_FW}).

  \textbf{First and second examples of FWs:}

\h Consider a couple of longitudinal intensity patterns, $|F(z)|^2$, in the range $0 \leq z \leq L$, given by a
sequence of two and three unitary step functions, described by

 \begin{equation} \label{2_p_intensidade_zeq}
                F(z)= \left\{ \begin{array}{rl}
                                        1, &  \text{for} ~~ l_{1} < z < l_{2}\\
                                        1, &  \text{for} ~~ l_{3} < z < l_{4}\\
                                        0, &  ~~~\text{elsewhere}
                \end{array}
                \right.
        \end{equation}
                \begin{equation} \label{3_p_intensidade_zeq}
                F(z)= \left\{ \begin{array}{rl}
                                        1, &  \text{for} ~~ l_{1} < z < l_{2}\\
                                        1, &  \text{for} ~~ l_{3} < z < l_{4}\\
                                        1, &  \text{for} ~~ l_{5} < z < l_{6}\\
                                        0, &  ~~~\text{elsewhere}
                \end{array}
                \right.
        \end{equation}

with $l_{1} = 10cm $, $l_{2} = 40cm$, $l_{3} = 50cm $, $l_{4} = 60cm$ in Eq.\ref{2_p_intensidade_zeq} and with
$l_{1} = 10cm $, $l_{2} = 20cm$, $l_{3} = 30cm $, $l_{4} = 50cm$, $l_{5} = 60cm $, $l_{6} = 70cm$ in
Eq.\ref{3_p_intensidade_zeq}.

\h As it was said, we use the solution (\ref{campo_FW}) with $k_{zn}$ and $A_n$ given by Eqs. (\ref{kz}) and
(\ref{An}). Now we have to choose the value of $Q$, which defines the spot size of the resulting FW
and, together with $L$, determines $N_{max}$. Moreover, from the experimental point of view, it is very
important to make an appropriated choice to $Q$ in order to respect the resolution limit of the SLM.
Concerning to this, and as a sufficient (but sometimes not necessary) condition, we should have the highest
value of $k_{\rho n}$, given by $k_{\rho n = - N}$, limited by the bandwidth magnitude of the SLM.

\h Using $Q = 0.9999943\,k$, we get $N_{max} = 9$ and a minimum diameter of $D=9.55$mm, which is
compatible with the dimensions of the SLM display. Finally, in this case we have $k_{\rho n = - N} = 4.74 \times
10^{4} m^{-1}$, which respect the bandwidth of SLM.

\h In this case we have $k_{\rho n=0} = 3.3525\times 10^{4} m^{-1}$, which implies in a spot of radius $\Delta
\rho_0 = 71.6\, \mu$m to the resulting FW.

\h Of course, the greater the number of terms in the series (\ref{campo_FW}), the better the result, i.e., the
resulting FW will be further closer to the desired beam.

\begin{figure}[H]
                \centering
                \includegraphics[scale=2.3]{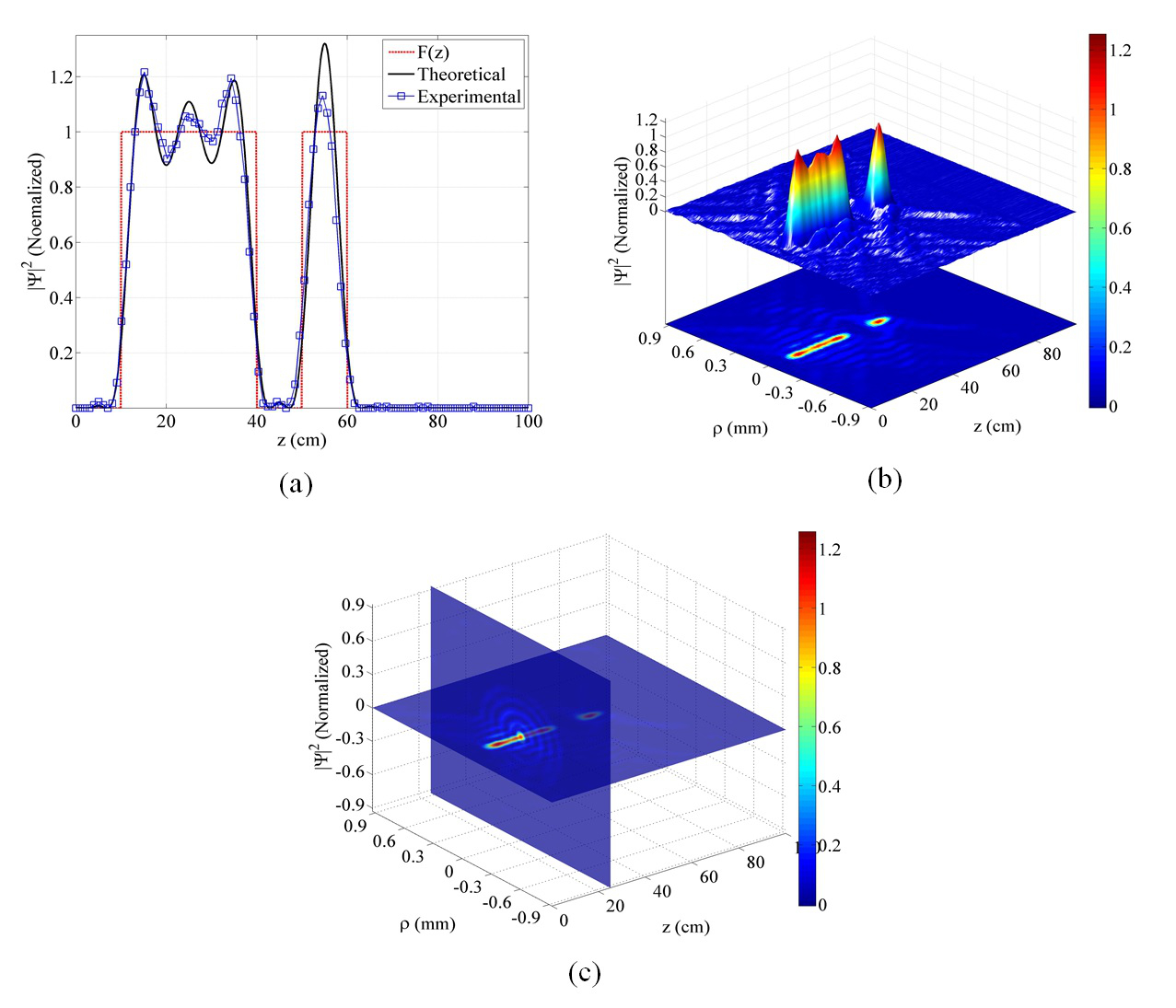}
                \caption{(a) Comparision among the longitudinal (on-axis) intensity patterns: $|F(z)|^2$, with $F(z)$
                given by Eq.\ref{2_p_intensidade_zeq}, and the respective theoretical and experimental
                FWs; (b) Three-dimensional and projected shapes of the experimental FW;
                (c) Intensity pattern slices of the experimental FW.}
        \label{2_degraus}
        \end{figure}

\begin{figure}[H]
                \centering
                \includegraphics[scale=2.3]{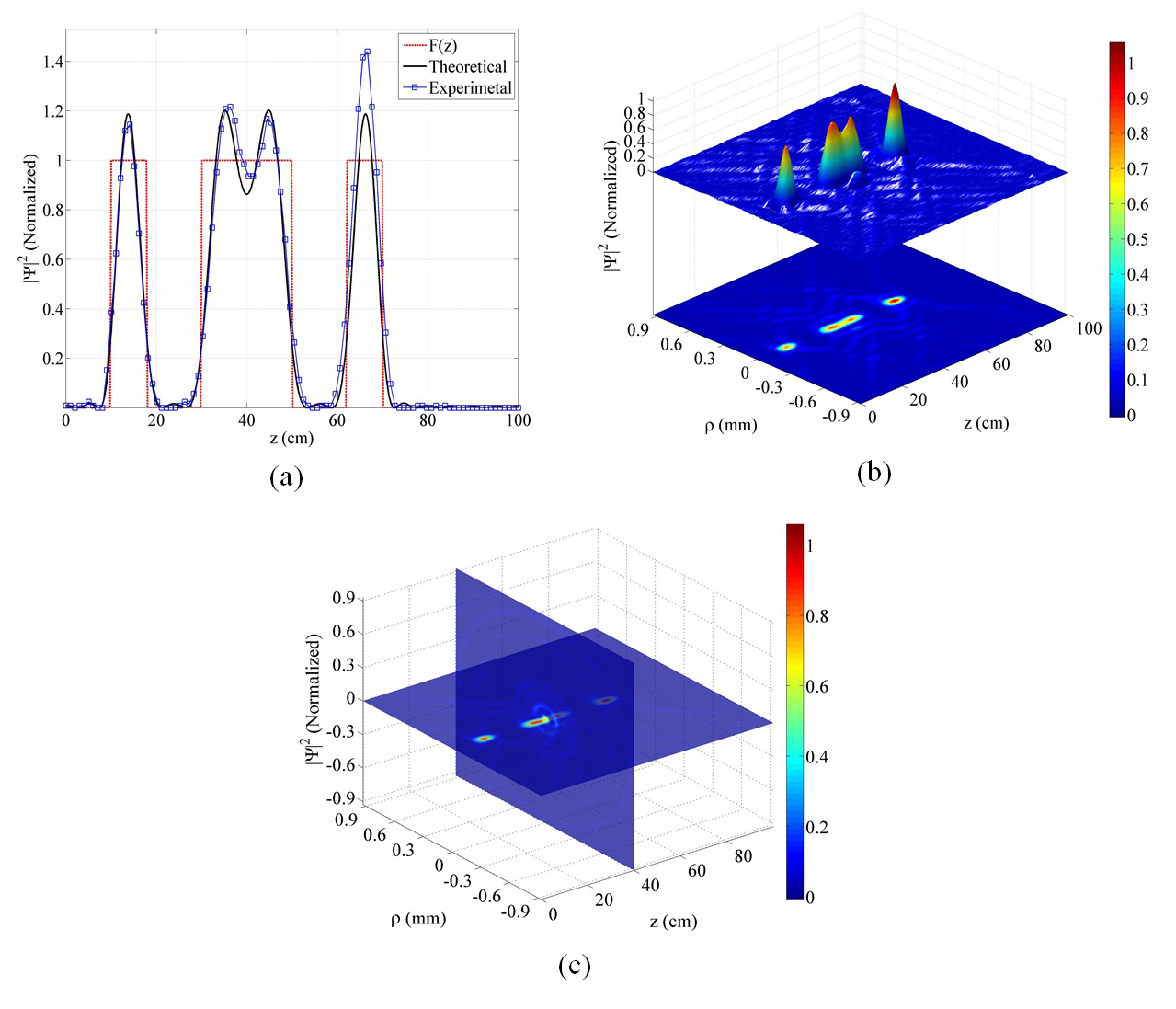}
                \caption{(a) Comparision among the longitudinal (on-axis) intensity patterns: $|F(z)|^2$, with $F(z)$ given
                by Eq.\ref{3_p_intensidade_zeq}, and the respective theoretical and experimental FWs;
                (b) Three-dimensional and projected shapes of the experimental FW;
                (c) Intensity pattern slices of the experimental FW.}
                \label{3_degraus}
        \end{figure}

\h The results, theoretical and experimental, are shown in Figures \ref{2_degraus} and \ref{3_degraus}.

\begin{figure}[H]
               \centering
                \includegraphics[scale=2]{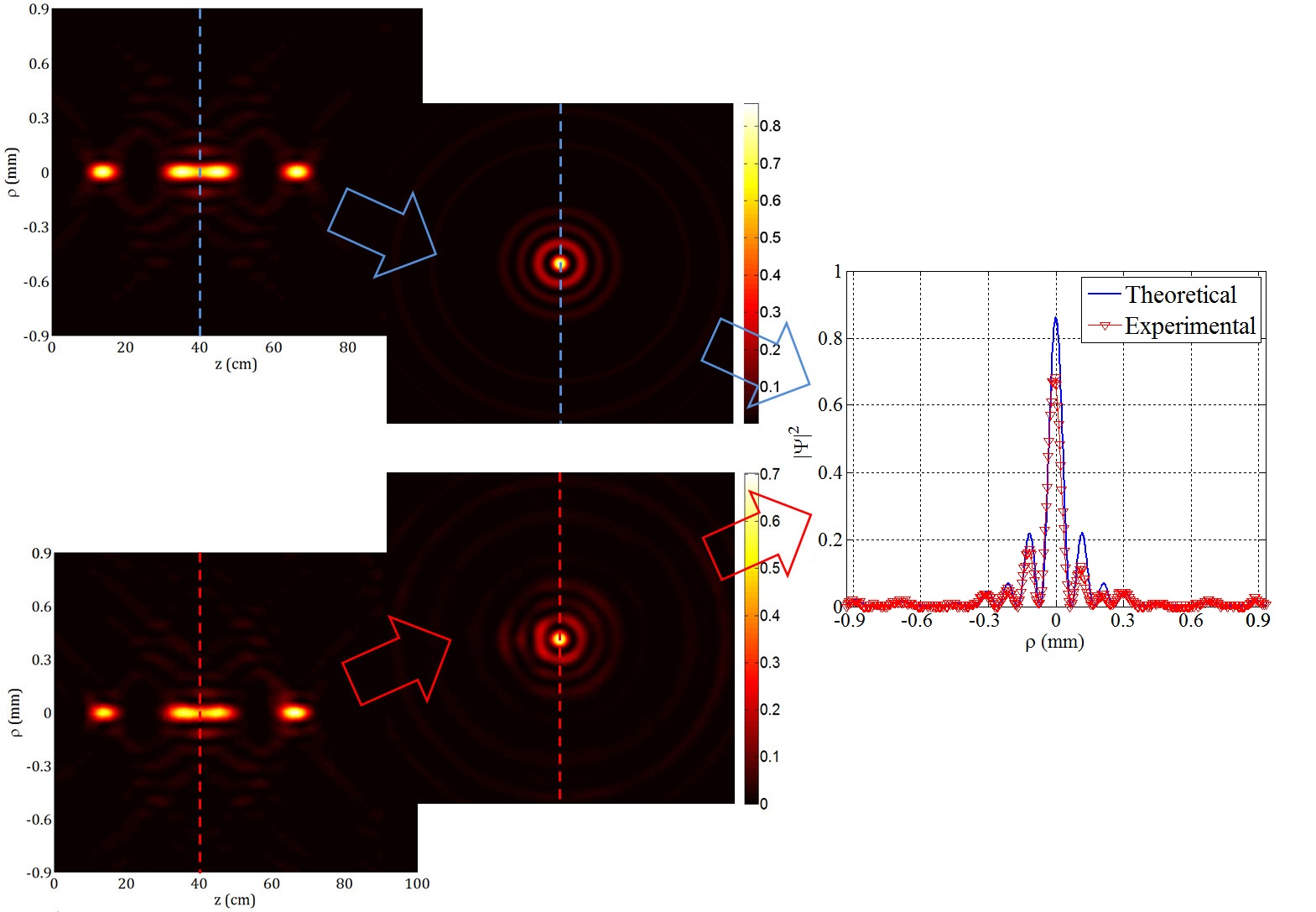}
                \caption{Comparison among the orthogonal projection and the transverse pattern of the intensities of the
                theoretical (up) and experimental (down) FWs with $F(z)$ given by Eq.(\ref{2_p_intensidade_zeq}).}
                \label{comparacao_3_degraus}
          \end{figure}

 \h Figure \ref{comparacao_3_degraus} shows the comparision among the longitudinal projection intensity and transversal
 pattern for $|F(z)|^2$, given by Eq.(\ref{3_p_intensidade_zeq}), these results theoretical and experimental FW are in
 excellent agreement.

It is easy to see that these FWs have nondiffracting properties, i.e., they resist to the diffraction effects
for long distances. For instance, in example 1 we see an intensity and spot size invariance over $35$ cm, while
a gaussian beam with the same initial spot radius (of $71\mu$m) doubles its transverse width after $2.5$ cm.

      \textbf{Third example:}

\h Here we choose a ladder-shaped longitudinal intensity pattern, $|F(z)|^2$, being

            \begin{equation} \label{intensidade_z_diferente_3eq}
                F(z)= \left\{ \begin{array}{rl}
                                        0.3, &  \text{for} ~~ l_{1} < z < l_{2}\\
                                        0.6, &  \text{for} ~~ l_{2} < z < l_{3}\\
                                        1   , &  \text{for} ~~ l_{3} < z < l_{4}\\
                                        0   , &  ~~~\text{elsewhere}
                \end{array}
                \right.
        \end{equation}
with $l_{1} = 5cm $, $l_{2} = 30cm$, $l_{3} = 50cm$, $l_{4} = 70cm$.

\begin{figure}[H]
                \centering
               \includegraphics[scale=2.3] {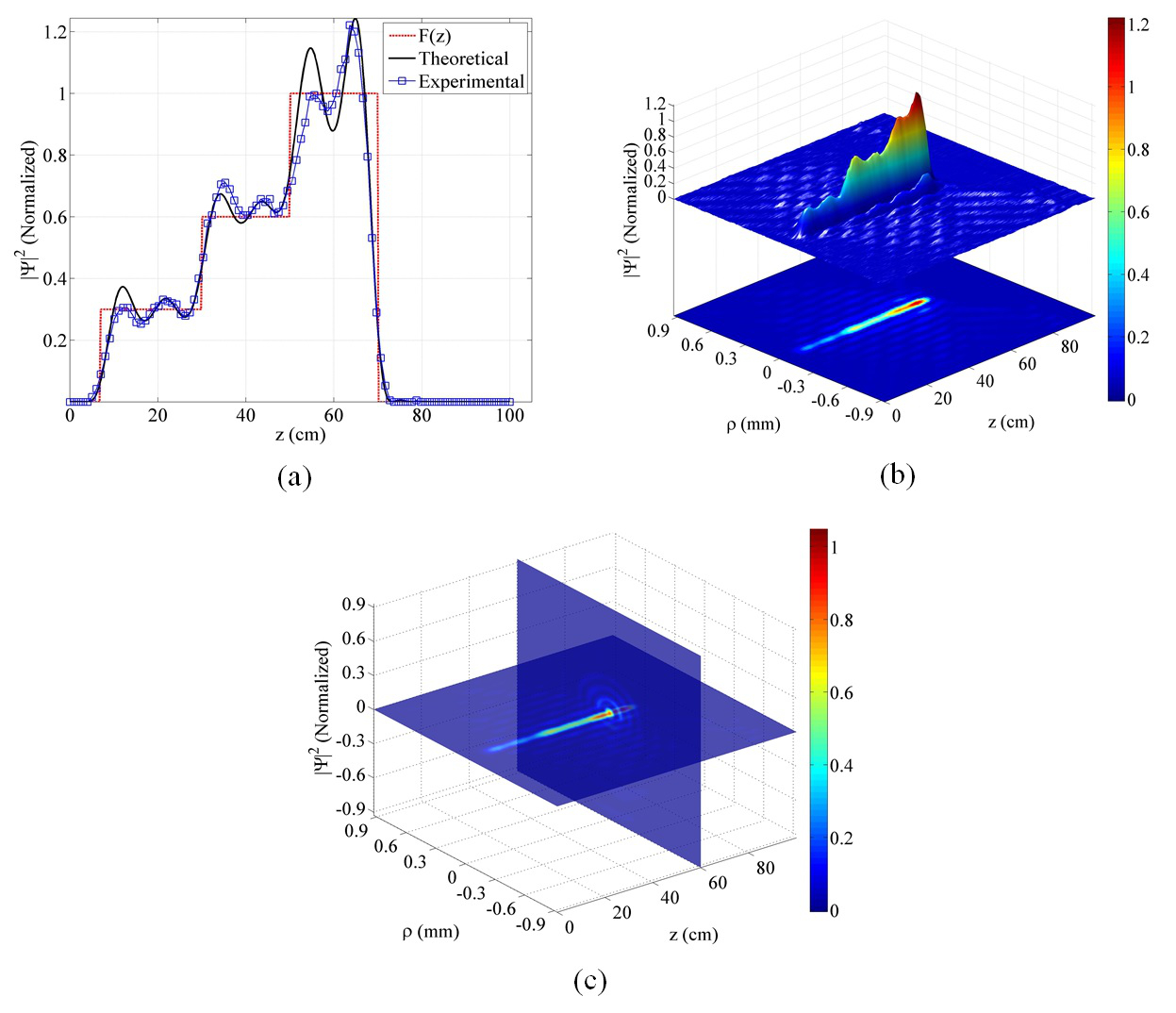} 
                \caption{(a) Comparison among the longitudinal (on-axis) intensity patterns: $|F(z)|^2$, with $F(z)$ given by
                Eq.\ref{intensidade_z_diferente_3eq}, and the respective theoretical and experimental FWs;
                (b) Three-dimensional and projected shapes of the experimental FW; (c) Intensity pattern slices of
                experimental the FW.}
                \label{escada}
        \end{figure}

\h Figure \ref{escada} show the theoretical and experimental Frozen Waves
obtained. We observe an excellent agreement between them.

            \textbf{Fourth example:}

\h Another interesting case is a swab-shaped longitudinal intensity pattern. This can be reached by choosing

        \begin{equation} \label{intensidade_z_diferente_2eq}
                F(z)= \left\{ \begin{array}{rl}
                                        1, &  \text{for} ~~ l_{1} < z < l_{2}\\
                                        1/2, &  \text{for} ~~ l_{2} < z < l_{3}\\
                                        1, &  \text{for} ~~ l_{3} < z < l_{4}\\
                                        0, &  ~~~\text{elsewhere}
                \end{array}
                \right.
        \end{equation}
whith, $l_{1} = 10$cm, $l_{2} = 20$cm, $l_{3} = 60$cm, $l_{4} = 70$cm.

\begin{figure}[H]
                \centering
                \includegraphics[scale=2.3] {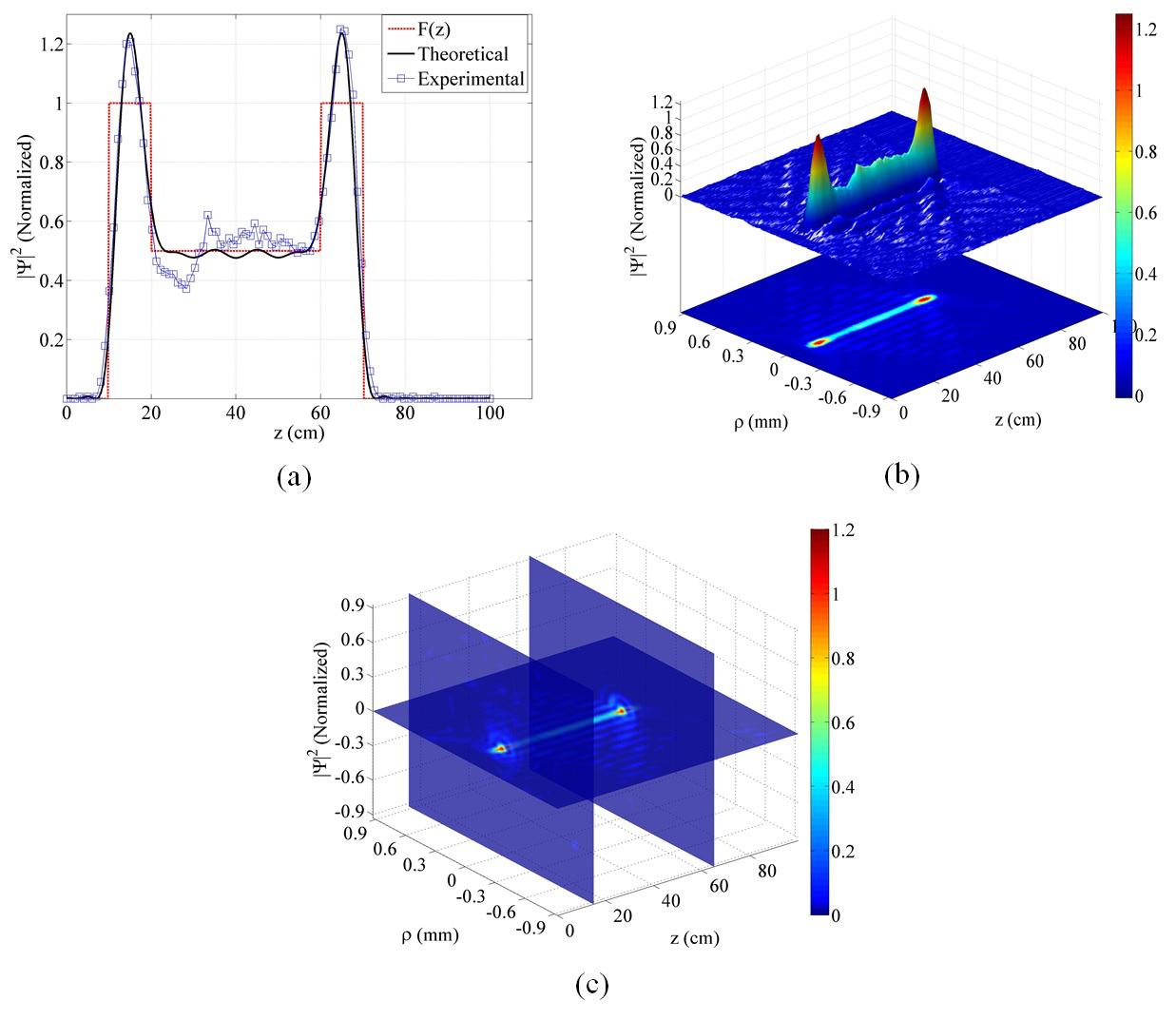} 
                \caption{(a) Comparision among the longitudinal (on-axis) intensity patterns: $|F(z)|^2$, with $F(z)$ given by
                Eq.\ref{intensidade_z_diferente_2eq}, and the respective theoretical and experimental FWs;
                (b) Three-dimensional and projected shapes of the experimental FW;
                (c) Intensity pattern slices of the experimental FW.}
                \label{cotonete}
       \end{figure}

\h Our results are shown in the Figure \ref{cotonete}. Again, there is an
excellent agreement between the theoretical FW ant that generated experimentally.

            \textbf{Fifth example:}

\h As we already said, it is possible to generate FWs with higher-order Bessel beams ($\nu>0$ in solution
(\ref{campo_FW})). In these cases, the desired longitudinal intensity pattern is shifted from the axis
($\rho=0$) to a cylindrical surface, whose radius may be approximately calculated using (\ref{raio cilindro}).
For details the reader is invited to consult the references
\cite{Zamboni;FW,Zamboni-Rached:05,Zamboni-Rached:06,Zamboni-Rached:10}.

\h In this example we choose a longitudinal intensity pattern given by an exponential function of increasing
intensity along the propagation direction, more specifically we choose:

        \begin{equation} \label{3d_teo_ordem_2eq}
                \centering
                F(z)= \left\{ \begin{array}{rl}
                                        \exp(qz);, &  \text{for} ~~ l_{1} < z < l_{2}\\
                                        0   , &  ~~~\text{elsewhere}
                \end{array}
                \right.
        \end{equation}
         with $q = 4/7$ ($l_{1} = 10cm $, $l_{2} = 50cm$)

\h Now, instead of using zero-order Bessel beams in the solution (\ref{campo_FW}), we will use $\nu=2$
(second-order Bessel beams). Remembering that we use the solution (\ref{campo_FW}) with $k_{zn}$ and $A_n$ given
by Eqs. (\ref{kz}) and (\ref{An}), and that (in the first five examples) we adopted $Q = 0.9999943\,k$, $N=9$
and $L=100\,$cm.

\begin{figure}[H]
                \centering
                \includegraphics[scale=2.3] {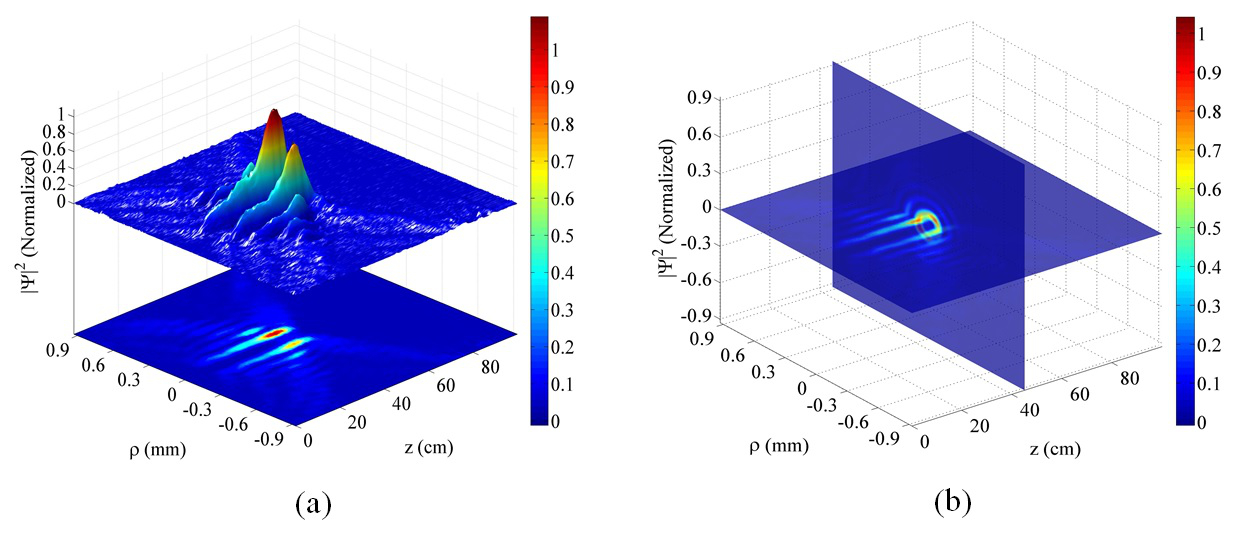}
                \caption{(a) Intensity pattern of experimental result of the chosen $|F(z)|^2$, with $F(z)$ given by
                Eq.(\ref{3d_teo_ordem_2eq});
                (b) Intensity pattern slices of the experimental FW.}
                \label{3d_com_projecao_ordem_2}
       \end{figure}

\h Figure \ref{3d_com_projecao_ordem_2} shows the 3D intensity of the experimental FW. We can easily see that the desired
intensity pattern occurs on a cylindrical surface.

There is an excellent agreement between theory and experiment.

   \textbf{Sixth example:}

Finally the last two examples was made in reflective SLM Setup 2, in this case due large bandwidth we choice $Q
= 0.99996\,k$, and from Eq.\ref{Nmax} $N_{max}=24.02$, if we choose $N = 24$, we find the maximum possible value
of the wave number transverse in Bessel beams superpositions $k_{\rho n=N} = 12.5\times 10^{4} m^{-1}$. In this
case we have $k_{\rho n=0} = 8.88\times 10^{4} m^{-1}$, which implies in a spot of radius $\Delta \rho_0 = 27\,
\mu$m to the resulting FW.

\begin{figure}[H]
                \centering
                \includegraphics[scale=2.3] {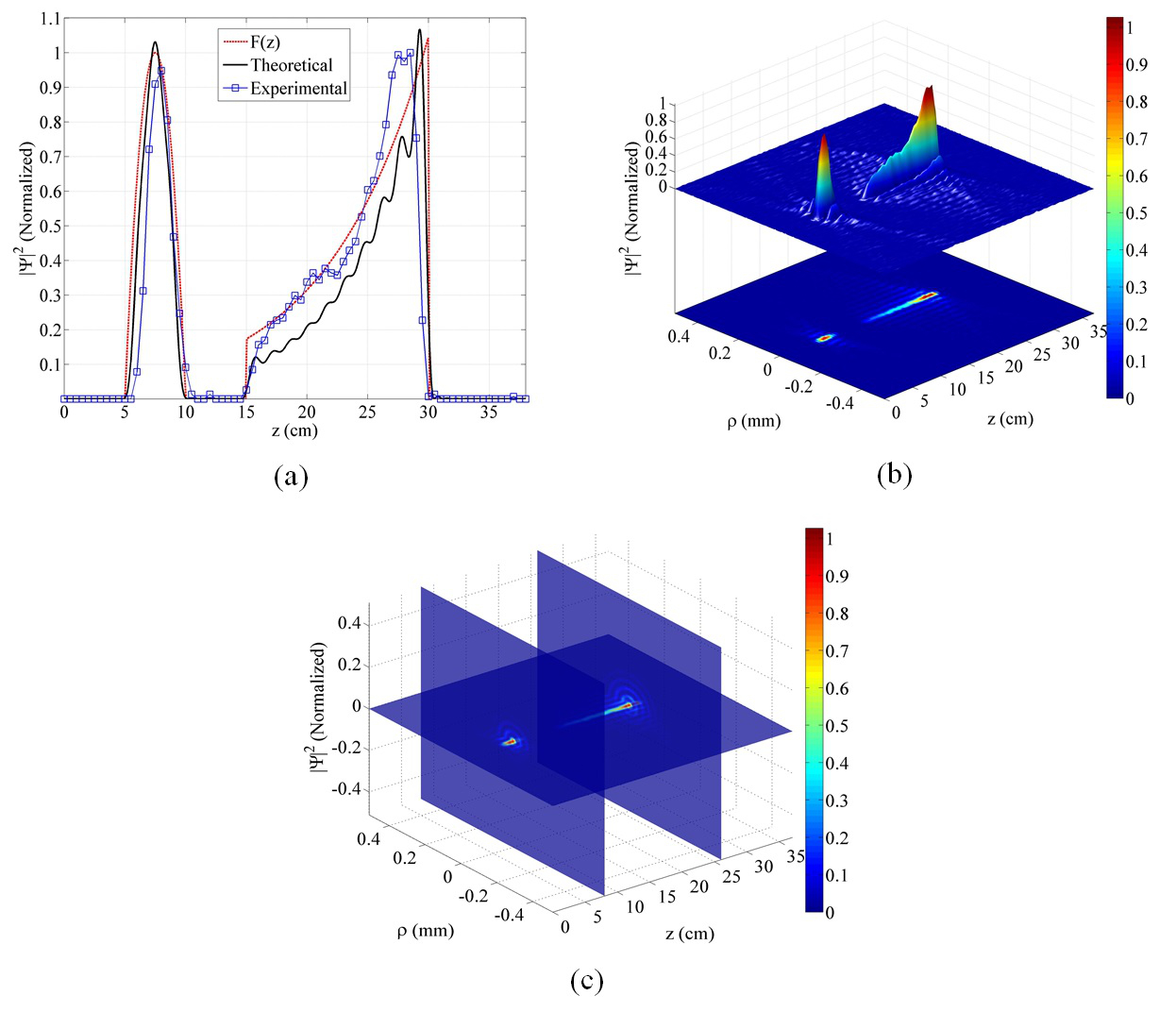} 
                \caption{(a) Comparison among the longitudinal (on-axis) intensity patterns: $|F(z)|^2$, with $F(z)$ given by
                Eq.\ref{1_p_expone_inten_zeq}, and the respective theoretical and experimental FWs;
                (b)  Three-dimensional and projected shapes of the experimental FW;
                (c) Intensity pattern slices of the experimental FW.}
                \label{x_quad_com_exp}
        \end{figure}

So for F(z) given by Eq.\ref{1_p_expone_inten_zeq},
        \begin{equation} \label{1_p_expone_inten_zeq}
                \centering
               F(z)= \left\{\begin{array}{rlr}
                                        \frac{-4(z^{2} - z(l_{2}+l_{1}) + l_{1}l_{2})}{(l_{2}-l_{1})^{2}};&  ~~~\text{for} ~~& l_{1} < z < l_{2}\\
                                                         \exp(qz)                                         ;&  ~~~\text{for} ~~& l_{3} < z < l_{4}\\
                                                              0                                  ;&   &~~\text{elsewhere}
                \end{array}
                \right.
        \end{equation}
        where $l_{1} = 5cm $, $l_{2} = 10cm$, $l_{3} = 15cm $, $l_{4} = 30cm$. We can see in Figure \ref{x_quad_com_exp}, for $F(z)$, the theoretical prediction and the experimental results.

\textbf{Seventh example:}

To generate a FW with the desired longitudinal intensity pattern concentrated over a cylindrical surface, we
choose $F(z)$ given by Eq.(\ref{ordem_3_inten_z_eq}), with $\nu=3$ in Eq.(\ref{campo_FW}), i.e. we use an higher
order Bessel beam superposition.

    \begin{equation} \label{ordem_3_inten_z_eq}
                F(z)= \left\{\begin{array}{rl}
                                        1, &  \text{for} ~~ l_{1} < z < l_{2}\\
                                        0, &  ~~~\text{elsewhere}
                \end{array}
                \right.
        \end{equation}
        where, $l_{1}=15cm$, $l_{2}=25cm$.

\begin{figure}[H]
                \centering
                \includegraphics[scale=2.3] {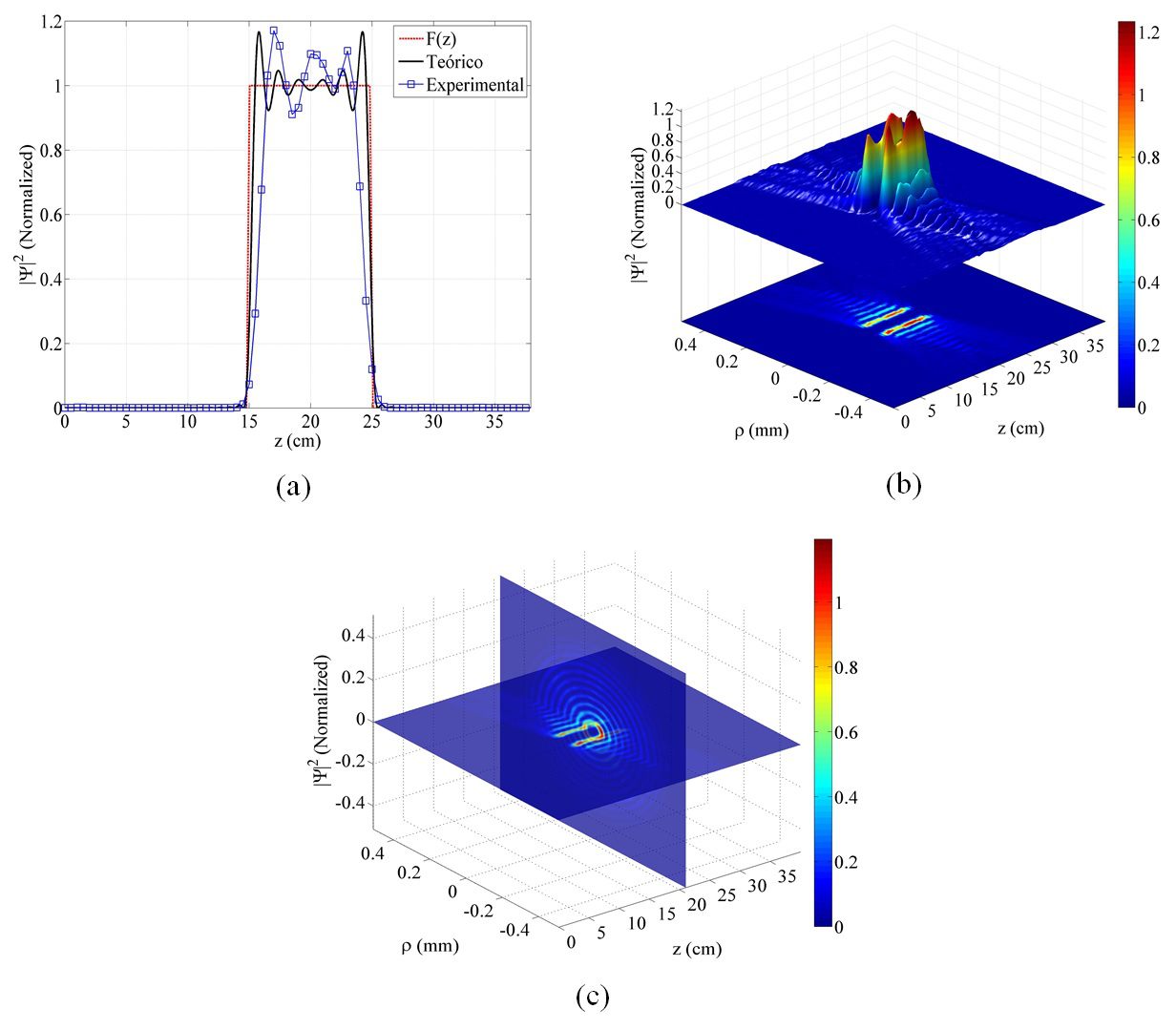} 
                \caption{(a) Comparison among the longitudinal (on-axis) intensity patterns: $|F(z)|^2$, with $F(z)$ given by
                Eq.\ref{ordem_3_inten_z_eq}, and the respective theoretical and experimental FWs;
                (b)  Three-dimensional and projected shapes of the experimental FW;
                (c) Intensity pattern slices of the experimental FW.}
                \label{ordem_3}
        \end{figure}

The results are show in Figure \ref{ordem_3}, this pattern can be used to many applications as wave or atoms
guide \cite{lan:2101,PhysRevLett.57.314,Song:99}.

\section{Conclusions}

The so called Frozen Waves are nondiffracting beams whose longitudinal (and, to lesser extent, also the
transverse) intensity pattern can be freely chosen a priori. In this paper, we present the experimental
generation of several Frozen Waves via computer generated holograms implemented in two types of spatial light
modulators, transmission and reflective. The experimental results for all FWs here considered are in excellent
agreement with theoretical predictions. This fact opens interesting possibilities for applying these and many
other FWs to scientific and technological purposes such as optical tweezers, remote sensing, atom guides,
optical scalpels or acoustic, electromagnetic or ultrasound of high intensity in medicine, among others.

\section{Acknowledgements}
The authors are grateful to Erasmo Recami and Mikiya Muramatsu for many stimulating contacts and discussions.

The authors acknowledge partial support from UFABC, UNICAMP, CAPES, FAPESP (grants 09/11429-2 and 11/51200-4)
and CNPQ (grants 307962/2010-5 and 309911/2011-7).

\end{document}